\documentclass{article}%
\usepackage{amsmath}
\usepackage{graphicx}%
\usepackage{amsfonts}%
\usepackage{amssymb}
\newtheorem{theorem}{Theorem}[section]

\newtheorem{corollary}[theorem]{Corollary}

\newtheorem{definition}[theorem]{Definition}

\newtheorem{remark}[theorem]{Remark}

\begin{document}

\title{On the Inertia of the Asynchronous Circuits}
\author{Serban. E. Vlad\\Oradea City Hall, Piata Unirii, Nr. 1, 3700, Oradea, Romania\\serban\_e\_vlad@yahoo.com, http://site.voila.fr/serban\_e\_vlad}
\date{}
\maketitle

\begin{abstract}
By making use of the notions and the notations from \cite{p3}, we present the
bounded delays, the absolute inertia and the relative inertia.

\end{abstract}

\section{Bounded Delays}

\begin{theorem}
The next system%
\begin{equation}
\underset{\xi\in\lbrack t-d_{r},t-d_{r}+m_{r}]}{\bigcap}u\left(  \xi\right)
\leq x\left(  t\right)  \leq\underset{\xi\in\lbrack t-d_{f},t-d_{f}+m_{f}%
]}{\bigcup}u\left(  \xi\right)  \label{e1}%
\end{equation}
where $u,x\in S$ and $0\leq m_{r}\leq d_{r},0\leq m_{f}\leq d_{f}$ defines a
DC if and only if%
\begin{equation}
d_{r}\geq d_{f}-m_{f},d_{f}\geq d_{r}-m_{r}\label{e2}%
\end{equation}
\textbf{Proof} The proof consists in showing that ($\ref{e2}$) implies for any
$u$ the existence of a solution $x$ of (\ref{e1}); any such $x$ satisfies
$x\in Sol_{SC}(u)$. If ($\ref{e2}$) is not fulfilled, it is proved that $u$
exists so that (\ref{e1}) has no solutions.
\end{theorem}

\begin{definition}
The system (\ref{e1}), when ($\ref{e2}$) is true, is called the
\textit{bounded delay condition} (BDC). $u,x$ are the \textit{input},
respectively the \textit{state} (or the \textit{output}); $m_{r},m_{f}$ are
the (\textit{rising,} f\textit{alling}) \textit{memories} (or
\textit{thresholds for cancellation}) and $d_{r},d_{f}$ , respectively
$d_{f}-m_{f},d_{r}-m_{r}$ are the (\textit{rising,} f\textit{alling})
\textit{upper bounds}, respectively the (\textit{rising,} f\textit{alling})
\textit{lower bounds of the transmission delay for transitions}. We say that
the tuple $(u,m_{r},d_{r},m_{f},d_{f})$ satisfies BDC. We shall also call
bounded delay condition the function $Sol_{BDC}^{m_{r},d_{r},m_{f},d_{f}%
}:S\rightarrow P^{\ast}(S)$ defined by%
\[
Sol_{BDC}^{m_{r},d_{r},m_{f},d_{f}}(u)=\{x|(u,m_{r},d_{r},m_{f},d_{f}%
)\ satisfies\ BDC\}
\]
\end{definition}

\begin{definition}
The inequalities ($\ref{e2}$) are called the \textit{consistency condition}
(CC) of BDC.
\end{definition}

\begin{theorem}
\label{t1.4}Let $0\leq m_{r}\leq d_{r},0\leq m_{f}\leq d_{f}$ and $0\leq
m_{r}^{^{\prime}}\leq d_{r}^{^{\prime}},0\leq m_{f}^{^{\prime}}\leq
d_{f}^{^{\prime}}$ so that CC is fulfilled for each of them.
\end{theorem}

\begin{description}
\item[a)] We note $d_{r}^{"}=\min(d_{r},d_{r}^{^{\prime}}),d_{f}^{"}%
=\min(d_{f},d_{f}^{^{\prime}}),m_{r}^{"}=d_{r}^{"}-\max(d_{r}-m_{r}%
,d_{r}^{^{\prime}}-m_{r}^{^{\prime}}),$ $m_{f}^{"}=d_{f}^{"}-\max(d_{f}%
-m_{f},d_{f}^{^{\prime}}-m_{f}^{^{\prime}})$. The next statements are equivalent:

\begin{enumerate}
\item[a.i)] $\forall u,Sol_{BDC}^{m_{r},d_{r},m_{f},d_{f}}(u)\wedge
Sol_{BDC}^{m_{r}^{^{\prime}},d_{r}^{^{\prime}},m_{f}^{^{\prime}}%
,d_{f}^{^{\prime}}}(u)\neq\emptyset$

\item[a.ii)] $d_{r}^{"}\geq d_{f}^{"}-m_{f}^{"},d_{f}^{"}\geq d_{r}^{"}%
-m_{r}^{"}$
\end{enumerate}

\item[ ] and if one of them is satisfied, then we have
\[
Sol_{BDC}^{m_{r},d_{r},m_{f},d_{f}}\wedge Sol_{BDC}^{m_{r}^{^{\prime}}%
,d_{r}^{^{\prime}},m_{f}^{^{\prime}},d_{f}^{^{\prime}}}=Sol_{BDC}^{m_{r}%
^{"},d_{r}^{"},m_{f}^{"},d_{f}^{"}}%
\]

\item[b)] We use the notations $d_{r}^{"}=\max(d_{r},d_{r}^{^{\prime}}%
),d_{f}^{"}=\max(d_{f},d_{f}^{^{\prime}}),$ $m_{r}^{"}=d_{r}^{"}-\min
(d_{r}-m_{r},d_{r}^{^{\prime}}-m_{r}^{^{\prime}}),$ $m_{f}^{"}=d_{f}^{"}%
-\min(d_{f}-m_{f},d_{f}^{^{\prime}}-m_{f}^{^{\prime}})$. The inequalities
$d_{r}^{"}\geq d_{f}^{"}-m_{f}^{"},d_{f}^{"}\geq d_{r}^{"}-m_{r}^{"}$ are
satisfied and%
\[
Sol_{BDC}^{m_{r},d_{r},m_{f},d_{f}}\vee Sol_{BDC}^{m_{r}^{^{\prime}}%
,d_{r}^{^{\prime}},m_{f}^{^{\prime}},d_{f}^{^{\prime}}}\subset Sol_{BDC}%
^{m_{r}^{"},d_{r}^{"},m_{f}^{"},d_{f}^{"}}%
\]
The previous inclusion becomes equality if and only if%
\[
\forall u,Sol_{BDC}^{m_{r},d_{r},m_{f},d_{f}}(u)\wedge Sol_{BDC}%
^{m_{r}^{^{\prime}},d_{r}^{^{\prime}},m_{f}^{^{\prime}},d_{f}^{^{\prime}}%
}(u)\neq\emptyset
\]

\item[c)] The next statements are equivalent:

\begin{enumerate}
\item[c.i)] $Sol_{BDC}^{m_{r},d_{r},m_{f},d_{f}}$ is deterministic

\item[c.ii)] The upper bounds and the lower bounds of the delays coincide:%
\[
d_{r}=d_{f}-m_{f},d_{f}=d_{r}-m_{r}%
\]

\item[c.iii)] The memories are null%
\[
m_{r}=m_{f}=0
\]

\item[c.iv)] The bounded delay degenerates in a translation%
\begin{equation}
\exists d\geq0,Sol_{BDC}^{m_{r},d_{r},m_{f},d_{f}}=I_{d} \label{c.iv}%
\end{equation}
\end{enumerate}

\item[d)] The next statements are equivalent

\begin{enumerate}
\item[d.i)] $Sol_{BDC}^{m_{r},d_{r},m_{f},d_{f}}\subset Sol_{BDC}%
^{m_{r}^{^{\prime}},d_{r}^{^{\prime}},m_{f}^{^{\prime}},d_{f}^{^{\prime}}}$

\item[d.ii)] $\ d_{r}^{^{\prime}}-m_{r}^{^{\prime}}\leq d_{r}-m_{r}\leq
d_{f}\leq d_{f}^{^{\prime}},\ d_{f}^{^{\prime}}-m_{f}^{^{\prime}}\leq
d_{f}-m_{f}\leq d_{r}\leq d_{r}^{^{\prime}}$
\end{enumerate}

\item[e)] $Sol_{BDC}^{m_{r},d_{r},m_{f},d_{f}}$ is time invariant

\item[f)] The next statements are equivalent

\begin{enumerate}
\item[f.i)] $Sol_{BDC}^{m_{r},d_{r},m_{f},d_{f}}$ is symmetrical

\item[f.ii)] $d_{r}=d_{f},m_{r}=m_{f}$
\end{enumerate}

\item[g)] $Sol_{BDC}^{m_{r}+m_{r}^{^{\prime}},d_{r}+d_{r}^{^{\prime}}%
,m_{f}+m_{f}^{^{\prime}},d_{f}+d_{f}^{^{\prime}}}$ is a BDC and we have%
\[
Sol_{BDC}^{m_{r}^{^{\prime}},d_{r}^{^{\prime}},m_{f}^{^{\prime}}%
,d_{f}^{^{\prime}}}\circ Sol_{BDC}^{m_{r},d_{r},m_{f},d_{f}}=Sol_{BDC}%
^{m_{r}+m_{r}^{^{\prime}},d_{r}+d_{r}^{^{\prime}},m_{f}+m_{f}^{^{\prime}%
},d_{f}+d_{f}^{^{\prime}}}%
\]
\end{description}

\section{Fixed and Inertial Delays}

\begin{definition}
\label{d2.1}Let $u,x\in S$ and $d\geq0$. The equation (see \ref{t1.4}
(\ref{c.iv}))%
\[
x\left(  t\right)  =u(t-d)
\]
is called the \textit{fixed delay condition} (FDC). The delay defined by this
equation is also called \textit{pure}, \textit{ideal} or \textit{non-inertial}%
. A delay different from FDC is called \textit{inertial}.
\end{definition}

\begin{corollary}
FDC is deterministic, time invariant, constant and symmetrical. The serial
connection of the FDC's coincides with the composition of the translations:%
\[
I_{d}\circ I_{d^{\prime}}=I_{d^{\prime}}\circ I_{d}=I_{d+d^{\prime}}%
,d\geq0,d^{\prime}\geq0
\]
\end{corollary}

\begin{remark}
At \ref{d2.1} inertia was defined to be the property of the DC's of being not
ideal. In particular the non-deterministic DC's, for example the non-trivial
BDC's (i.e. the BDC's with memory $m_{r}+m_{f}\gtrdot0$) are inertial.
\end{remark}

\section{Absolute Inertia}

\begin{definition}
The property%
\begin{align*}
\overline{x(t-0)}\cdot x(t)  &  \leq\underset{\xi\in\lbrack t,t+\delta_{r}%
]}{\bigcap}x\left(  \xi\right) \\
x(t-0)\cdot\overline{x(t)}  &  \leq\underset{\xi\in\lbrack t,t+\delta_{f}%
]}{\bigcap}\overline{x\left(  \xi\right)  }%
\end{align*}
true for $\delta_{r}\geq0,\delta_{f}\geq0$ is called the \textit{absolute
inertial condition} (AIC), or the non-zenoness condition. $\delta_{r}%
,\delta_{f}$ are called \textit{inertial parameters}. If it is fulfilled, we
say that the tuple $(\delta_{r},\delta_{f},x)$ satisfies AIC. We also call AIC
the set $Sol_{AIC}^{\delta_{r},\delta_{f}}\subset S$ defined by%
\[
Sol_{AIC}^{\delta_{r},\delta_{f}}=\{x|(\delta_{r},\delta_{f}%
,x)\ satisfies\ AIC\}
\]
\end{definition}

\begin{remark}
AIC means that if $x$ switches from $0$ to $1$, then it remains $1$ at least
$\delta_{r}\geq0$ time units + the dual property. To be remarked the trivial
situation $\delta_{r}=\delta_{f}=0$.
\end{remark}

\begin{definition}
Let $i$ a DC satisfying $\forall u,i(u)\wedge Sol_{AIC}^{\delta_{r},\delta
_{f}}\neq\emptyset$. The DC $i\wedge Sol_{AIC}^{\delta_{r},\delta_{f}}$ is
called \textit{absolute inertial delay condition} (AIDC). $Sol_{BDC}%
^{m_{r},d_{r},m_{f},d_{f}}\wedge Sol_{AIC}^{\delta_{r},\delta_{f}} $ is called
\textit{bounded absolute inertial delay condition} (BAIDC).
\end{definition}

\begin{theorem}
The numbers $0\leq m_{r}\leq d_{r},0\leq m_{f}\leq d_{f}$ \ with CC true and
$\delta_{r}\geq0,\delta_{f}\geq0$ are given. The next statements are equivalent:
\end{theorem}

\begin{description}
\item[a)] $\forall u,Sol_{BDC}^{m_{r},d_{r},m_{f},d_{f}}(u)\wedge
Sol_{AIC}^{\delta_{r},\delta_{f}}\neq\emptyset$

\item[b)] $\delta_{r}+\delta_{f}\leq m_{r}+m_{f}$
\end{description}

\begin{corollary}
$0\leq m_{r}\leq d_{r},0\leq m_{f}\leq d_{f}$, $0\leq m_{r}^{^{\prime}}\leq
d_{r}^{^{\prime}},0\leq m_{f}^{^{\prime}}\leq d_{f}^{^{\prime}}$ and
$\delta_{r}\geq0,\delta_{f}\geq0,\delta_{r}^{^{\prime}}\geq0,\delta
_{f}^{^{\prime}}\geq0$ satisfy $d_{r}\geq d_{f}-m_{f},d_{f}\geq d_{r}%
-m_{r},d_{r}^{^{\prime}}\geq d_{f}^{^{\prime}}-m_{f}^{^{\prime}}%
,d_{f}^{^{\prime}}\geq d_{r}^{^{\prime}}-m_{r}^{^{\prime}},$ $\delta
_{r}+\delta_{f}\leq m_{r}+m_{f},\delta_{r}^{^{\prime}}+\delta_{f}^{^{\prime}%
}\leq m_{r}^{^{\prime}}+m_{f}^{^{\prime}}$. In such conditions $Sol_{BDC}%
^{m_{r},d_{r},m_{f},d_{f}}\wedge Sol_{AIC}^{\delta_{r},\delta_{f}}$ ,
$Sol_{BDC}^{m_{r}^{^{\prime}},d_{r}^{^{\prime}},m_{f}^{^{\prime}}%
,d_{f}^{^{\prime}}}\wedge Sol_{AIC}^{\delta_{r}^{^{\prime}},\delta
_{f}^{^{\prime}}},Sol_{BDC}^{m_{r}+m_{r}^{^{\prime}},d_{r}+d_{r}^{^{\prime}%
},m_{f}+m_{f}^{^{\prime}},d_{f}+d_{f}^{^{\prime}}}\wedge Sol_{AIC}^{\delta
_{r}^{^{\prime}}\delta_{f}^{^{\prime}}}$ are BAIDC's and the next property of
the serial connection holds:%
\[
(Sol_{BDC}^{m_{r}^{^{\prime}},d_{r}^{^{\prime}},m_{f}^{^{\prime}}%
,d_{f}^{^{\prime}}}\wedge Sol_{AIC}^{\delta_{r}^{^{\prime}}\delta
_{f}^{^{\prime}}})\circ(Sol_{BDC}^{m_{r},d_{r},m_{f},d_{f}}\wedge
Sol_{AIC}^{\delta_{r},\delta_{f}})\subset
\]%
\[
\subset Sol_{BDC}^{m_{r}+m_{r}^{^{\prime}},d_{r}+d_{r}^{^{\prime}},m_{f}%
+m_{f}^{^{\prime}},d_{f}+d_{f}^{^{\prime}}}\wedge Sol_{AIC}^{\delta
_{r}^{^{\prime}}\delta_{f}^{^{\prime}}}%
\]
\end{corollary}

\section{Relative Inertia}

\begin{definition}
$0\leq\mu_{r}\leq\delta_{r},0\leq\mu_{f}\leq\delta_{f}$ and $u,x\in S $ are
given. The property%
\begin{align*}
\overline{x(t-0)}\cdot x(t)  &  \leq\underset{\xi\in\lbrack t-\delta
_{r},t-\delta_{r}+\mu_{r}]}{\bigcap}u\left(  \xi\right) \\
x(t-0)\cdot\overline{x(t)}  &  \leq\underset{\xi\in\lbrack t-\delta
_{f},t-\delta_{f}+\mu_{f}]}{\bigcap}\overline{u\left(  \xi\right)  }%
\end{align*}
is called the \textit{relative inertial condition} (RIC). $\mu_{r},\delta
_{r},\mu_{f},\delta_{f}$ are called \textit{inertial parameters}. If it is
fulfilled, we say that the tuple $(u,\mu_{r},\delta_{r},\mu_{f},\delta_{f},x)$
satisfies RIC. We also call RIC the function $Sol_{RIC}^{\mu_{r},\delta
_{r},\mu_{f},\delta_{f}}:S\rightarrow P^{\ast}(S)$ defined by%
\[
Sol_{RIC}^{\mu_{r},\delta_{r},\mu_{f},\delta_{f}}\left(  u\right)
=\{x|(u,\mu_{r},\delta_{r},\mu_{f},\delta_{f},x)\ satisfies\ RIC\}
\]
\end{definition}

\begin{theorem}
\label{t4.2}Let $0\leq\mu_{r}\leq\delta_{r},0\leq\mu_{f}\leq\delta_{f},u\in S$
and $x\in Sol_{RIC}^{\mu_{r},\delta_{r},\mu_{f},\delta_{f}}\left(  u\right)  $
arbitrary. If $\delta_{r}\geq\delta_{f}-\mu_{f},\delta_{f}\geq\delta_{r}%
-\mu_{r}$ then $x\in Sol_{AIC}^{\delta_{f}-\delta_{r}+\mu_{r},\delta
_{r}-\delta_{f}+\mu_{f}}$.
\end{theorem}

\begin{remark}
RIC states that the inertial delays 'model the fact that the practical
circuits will not respond (at the output) to two transitions (at the input)
which are very close together' \cite{p1}, \cite{p2}. Theorem \ref{t4.2}
connecting AIC and RIC makes use of the condition $\delta_{r}\geq\delta
_{f}-\mu_{f},\delta_{f}\geq\delta_{r}-\mu_{r}$ that is very similar to CC, but
with a different meaning.
\end{remark}

\begin{definition}
Let $i$ a DC with $\forall u,i\left(  u\right)  \wedge Sol_{RIC}^{\mu
_{r},\delta_{r},\mu_{f},\delta_{f}}\left(  u\right)  \neq\emptyset$. Then the
DC $i\wedge Sol_{RIC}^{\mu_{r},\delta_{r},\mu_{f},\delta_{f}}$ (see Theorem
4.4 c) in \cite{p3}) is called \textit{relative inertial delay condition}
(RIDC). In particular $Sol_{BDC}^{m_{r},d_{r},m_{f},d_{f}}\wedge
Sol_{RIC}^{\mu_{r},\delta_{r},\mu_{f},\delta_{f}}$ is called \textit{bounded
relative inertial delay condition} (BRIDC).
\end{definition}

\begin{theorem}
\label{t4.5}Let the numbers $0\leq m_{r}\leq d_{r},0\leq m_{f}\leq d_{f}$ .
The next conditions are equivalent
\end{theorem}

\begin{enumerate}
\item[a)] $\forall u,Sol_{BDC}^{m_{r},d_{r},m_{f},d_{f}}\left(  u\right)
\wedge Sol_{RIC}^{\mu_{r},\delta_{r},\mu_{f},\delta_{f}}\left(  u\right)
\neq\emptyset$

\item[b)] One of the next conditions is true

\begin{description}
\item[b.i)] $d_{f}-m_{f}\leq\delta_{r}\leq d_{r}\leq\delta_{r}-\mu_{r}%
+m_{r},d_{r}-m_{r}\leq\delta_{f}\leq d_{f}\leq\delta_{f}-\mu_{f}+m_{f}$

\item[b.ii)] $d_{r}-m_{r}+\mu_{r}\leq\delta_{r}\leq d_{f}-m_{f}\leq
d_{r},d_{f}-m_{f}+\mu_{f}\leq\delta_{f}\leq d_{r}-m_{r}\leq d_{f}$

\item[b.iii)] $d_{f}-m_{f}\leq\delta_{r}\leq d_{r}-m_{r}+\mu_{r}\leq
d_{r},d_{r}-m_{r}\leq\delta_{f}\leq d_{f}-m_{f}+\mu_{f}\leq d_{f}$

\item[b.iv)] $\delta_{r}\leq d_{f}-m_{f}\leq\delta_{r}+m_{r}-\mu_{r}\leq
d_{r},\delta_{f}\leq d_{r}-m_{r}\leq\delta_{f}+m_{f}-\mu_{f}\leq d_{f}$
\end{description}
\end{enumerate}

\begin{remark}
The equivalent conditions from Theorem \ref{t4.5} are of consistency of BRIDC,
they are stronger than CC (of BDC) and weaker than (see the hypothesis
$\delta_{r}\geq\delta_{f}-\mu_{f},\delta_{f}\geq\delta_{r}-\mu_{r}$ from
Theorem \ref{t4.2})
\begin{align*}
d_{f}-m_{f}  &  \leq\delta_{f}-\mu_{f}\leq\delta_{r}\leq d_{r}\\
d_{r}-m_{r}  &  \leq\delta_{r}-\mu_{r}\leq\delta_{f}\leq d_{f}%
\end{align*}
\end{remark}

\begin{theorem}
\label{t4.7}Let $0\leq m_{r}\leq d_{r},0\leq m_{f}\leq d_{f}$ so that CC is
fulfilled and $u\in S$ arbitrary. The next statements are equivalent:
\end{theorem}

\begin{enumerate}
\item[a)] $x\in Sol_{BDC}^{m_{r},d_{r},m_{f},d_{f}}\left(  u\right)  \wedge
Sol_{RIC}^{m_{r},d_{r},m_{f},d_{f}}\left(  u\right)  $

\item[b)]
\begin{align*}
\overline{x(t-0)}\cdot x(t)  &  =\overline{x(t-0)}\cdot\underset{\xi\in\lbrack
t-d_{r},t-d_{r}+m_{r}]}{\bigcap}u\left(  \xi\right) \\
x(t-0)\cdot\overline{x(t)}  &  =x(t-0)\cdot\underset{\xi\in\lbrack
t-d_{f},t-d_{f}+m_{f}]}{\bigcap}\overline{u\left(  \xi\right)  }%
\end{align*}
\end{enumerate}

\begin{theorem}
Any of the previous equivalent conditions defines a deterministic, time
invariant, constant DC.
\end{theorem}

\begin{remark}
The deterministic situation \ref{t4.7} of BRIDC has as special case $I_{d}$,
happening when $m_{r}=m_{f}=0,d_{r}=d_{f}=d$. On the other hand the serial
connection of the BRIDC's is not a BRIDC. We also mention the possibility of
replacing the functions $\underset{\xi\in\lbrack t-d_{r},t-d_{r}+m_{r}%
]}{\bigcap}u\left(  \xi\right)  ,\underset{\xi\in\lbrack t-d_{f},t-d_{f}%
+m_{f}]}{\bigcup}u\left(  \xi\right)  $ with $\underset{\xi\in\lbrack
t-d_{r},t)}{\bigcap}u\left(  \xi\right)  ,\underset{\xi\in\lbrack t-d_{f}%
,t)}{\bigcup}u\left(  \xi\right)  $ in BDC, the functions $\underset{\xi
\in\lbrack t,t+\delta_{r}]}{\bigcap}x\left(  \xi\right)  ,\underset{\xi
\in\lbrack t,t+\delta_{f}]}{\bigcap}\overline{x\left(  \xi\right)  }$ with
$\underset{\xi\in\lbrack t,t+\delta_{r})}{\bigcap}x\left(  \xi\right)
,\underset{\xi\in\lbrack t,t+\delta_{f})}{\bigcap}\overline{x\left(
\xi\right)  }$ in AIC, the functions $\underset{\xi\in\lbrack t-\delta
_{r},t-\delta_{r}+\mu_{r}]}{\bigcap}u\left(  \xi\right)  $ and $\underset
{\xi\in\lbrack t-\delta_{f},t-\delta_{f}+\mu_{f}]}{\bigcap}\overline{u\left(
\xi\right)  }$ with $\underset{\xi\in\lbrack t-\delta_{r},t)}{\bigcap}u\left(
\xi\right)  ,\underset{\xi\in\lbrack t-\delta_{f},t)}{\bigcap}\overline
{u\left(  \xi\right)  }$ in RIC etc. and some variants of the previous
definitions result. The last six functions are not signals.
\end{remark}

\end{document}